# On the astronomical content of the sacred landscape of Cusco in Inka times.

**Giulio Magli**
**Dipartimento di Matematica del Politecnico di Milano**
**P.le Leonardo da Vinci 32, 20133 Milano, Italy.**

## 1. Introduction: the Inka sacred landscape

It is very well known that the Inka landscape was a *sacred space* or *sacred landscape*.[1] The Inka world was in fact plenty of "holy things" – called *huacas* - such as mountains, springs, rocks, and shrines, and many aspects of this deep relationship between people and nature actually survive also nowadays.
The Inka perception of the landscape was incorporated in the structure of the capital of the empire, Cusco. Founded (or most probably re-founded) around the 12 century a.C., the city lies at the confluence of two rivers and, at least according to a long-standing tradition, was conceived in the form of a puma. The hill surmounted by the huge walls of the Sacsahuaman form the head, the large square Huacaypata lies between the legs, and the main temple, the Coricancha, lies in the genitals. The tail of the Puma or *pumachupan* is formed at the confluence of the Tullumayo and the Huatanay rivers in the Vilcanota river (Fig. 1).
According to many of the accounts of Indian life and lore written in Spanish shortly after the conquest , the city was planned in such a way that the main temple was sitting at the confluence of 42 lines, called *ceques*, each containing a varying number of huacas. The total number of huacas was 328. The ceques lines were grouped into four radial quarters, each one corresponding to one of the four "parts" of the state. The Inka state or *Tahuantinsuyu* was in fact thought of as made of four parts, *Chinchaysuyu* and *Antisuyu*, associated with *hanan* ("upper"), and *Collasuyu* and *Cuntisuyu*, associated with *hurin* (below).
Much effort has been devoted to the reconstruction of the system of the ceque lines, an effort which of course is devoted towards the understanding of the meaning of this complex "cosmography" as well. Tom Zuidema, the pioneer in this field, was the first to recognize the relevance of astronomy in the structure of the system.[2] The ceque system as a whole is in fact connected with the Inka lunar-stellar calendar, and several of the orientations of the lines have an astronomical origin and mark relevant astronomical phenomena at the horizon, such as the rising of the sun at the solstices.[3]
In recent years, Brian Bauer completed a wide fieldwork aimed to the identification of the *huacas* of the system, and succeeded in identifying more than a half of them (Fig. 2).[4] I will not, by no means, attempt a review of Bauer's work, and I will only resume some relevant facts emerging from it which will be of importance in what follows:

1) The ceque lines are not strictly straight: the sub-lines connecting huacas can deviate each other within the same ceque. However, almost all lines do not intersect each other.
2) Altough not all the lines start exactly from the Coricancha, the radial disposition of the structure is clearly confirmed.
3) Huacas consist of various typologies. Among them: spring or sources of water (29%) , standing stones (29%), hills/mountain passes (10%) buildings (royal and temples) (9%) fields and flat places (9%), a few examples of other typologies such as tombs, caves, stone seats.
4) A few huacas were movable (e.g. movable stones).
5) The system was not completely fixed, that is, some huacas could be added or removed and some ceques could be changed. However, it is clear that the plan was conceived on the basis of a global project.

Today we have a reasonably clear picture of the role of the ceque system in the social organization of the capital (for instance, the lines were divided in clusters of three types, reflecting the importance of the huacas and of the families who were in charge of keeping it and making offers to the shrines). The same cannot be said, however, about the origin and the planning of the system. It is, in fact, pretty clear that a complex symbolic structure like this did not form spontaneously: it involved a huge planning fieldwork, during which some person was in charge of deciding the number of lines, the number of huacas, the type of huacas, and if a huaca had to be assigned to a ceque or to another.
The aim of the present paper is to propose possible unnoticed connections between the planning of the ceque system and the lore - in particular, the sky-lore - of the Inkas.

## 2. The *Khipus* and the sacred landscape

It was noted already by the first cronachers that the radial organization of the sacred space of Cusco corresponds to the "radial way of thinking" of the Inka written records. These records, called *Khipus* , consist in strings of different origin (e.g. cotton or fibres) and colours, attached to a "master" string. Each string carried knots, which could be of various types, and the knots could be clustered in groups.
There is no doubt that these devices were used *at least* to keep track of data in order to help the memory of the maker. For instance, if the maker of the Khipu was an exactor of taxes, he used it to record the type, the quantity, and the place where tributes were collected (numbers were annotated on a decimal basis using the "hierarchy" of the nodes to indicate units, tens and so on). This fact originated the wide spread point of view that Khipus were *personal* writings, that is, the idea that a Khipu was only a "help to the memory" thus readable only by its maker. This idea conflicts with many chronicles which, more or less clearly, state that the Khipus were a form of writing and, much more important, does not make justice to the Inka intelligence.
In recent years, the idea that the Khipus were only personal writings has finally been shown to be wrong by Gary Urton.[5]
Basing on written chronicles in which the content of some Khipus is described, Urton showed that it was certainly possible to store *at least* statements of informative content in a complete way (such as "the village X gave Y Indians to attend to the mummy of Z"). According to Urton, who takes into account type of material, types of knots, spin of the fibres and so on, the "bytes" of information available should have reached the number of 1536, a quantity comparable with the number of signs of ideographic writings of other cultures. Further to Urton's work, the recent controversial discovery of the so called *Miccinelli Manuscripts* has opened the way to the possible existence of a different kind of Khipus, which were used as a support for written language.[6] Such special Khipus were very simple: to each string a "keyword" in the form of a painted cotton banner was attached. The number of knots was used to tell to the reader which syllab of the keyword was to be extracted. Reading together the syllabs one obtained the words of the text (it should be stressed that the authenticity of the Miccinelli manuscripts is still debated however). The existence of syllabic Khipus would explain, by the way, the fact that the Khipus were persecuted by the conquerors and, in 1580, they were declared objects of idolatry and hunted for destruction, exactly as the maya codex's. Today, four maya codex's and some hundreds of Khipus survive.

The Khipus were thus certainly used at least to transmit quantitative information to all people able to read. Khipus were without any doubt sufficient to record, for instance, simple statements of historical, religious, or scientific nature, such as astronomical observations. For instance, in the *Nuova Cronica y Buen Gobierno* by Guaman Poma de Ayala the Inka astrologer, "who studies the sun, the moon, and all other heavenly bodies in order to know when to plant the fields", is represented as an old man carrying a Khipu (Fig. 3).

It is reasonable to think that the structure of the ceque system was originally recorded in a Khipu, but I would like to propose here the possibility of a further relationship between the radial ceque system and the radial Khipu notation. My proposal is that the Cusco ceque system might have been planned on the basis of a chosen Khipu. In other words, the visual analogy between the ceque system and "a" (generic) Khipu, might instead be an identification with a Khipu carrying specific information. For instance, taking into account the connection between the ceque lines and the calendar discovered by Zuidema, such information could be of astronomical-calendaring type. The idea is, that the different typologies of the huacas, which seem to vary from each other in a random way, could instead correspond to different typologies of knots and to different colours of strings, in such a way that the global project of the ceque system can be read as a monumental Khipu. If the existence of syllabic Khipus will be confirmed than the first huaca of each ceque could represent the keyword attached to the ceque-string (actually one of the syllabic keyword cited in the list given in the Miccinelli Manuscripts is "Puma").

A possibility to check if this idea can be reliable would be to construct "tree diagrams" for the possible continuation of ceque lines between different adjacent huacas as a function of the type and function of the huacas, trying to understand the reasons of choosing the next huacas for that specific ceque. In my opinion, this study can be of help, in any case, in order to understand the way in which the planning of the ceque system was conceived also in view of a further possible connection with astronomy, to be discussed in next section.

## 3. The dark cloud constellations and the sacred landscape

The Milky Way – called *Mayu* - was a central object in Inka astronomy and was individuated as a celestial counterpart of the Vilcanota river. The identification was so deep that actually the water flowing on the earth was thought of as the same water flowing in the celestial river and coming back to the earth in the rainy season. In many of the chronicles such as the famous one by Garcilaso de La Vega, it is reported that the Inka identified animals in the sky in the region of the Milky Way. Up to the seventies, however, the identification of what people believed to be Inka constellations in the sense *we* give to such a word, i.e. patterns formed in the sky connecting stars with lines, was not successful. In the meanwhile, it seems that nobody was giving the right credit to the chroniclers when the say that the Inka were actually viewing *dark* animals in the sky. Finally, the fundamental fieldwork carried out by Gary Urton with informants of the Quechua villages of Sonqo and (especially) Misminay, some 50 kilometres from the Inka capital Cusco, solved the enigma.[7]

The animals in the sky are not patterns formed connecting stars, but black regions of the Milky Way (dark clouds of interstellar matter, from the astronomer's viewpoint) whose contours are identified with contours of animals. Urton has been able to identify unambiguously the following dark cloud constellations (Fig. 4):

1) Serpent, between the star Adhara, in Canis Major, and the Southern Cross
2) Toad, near Southern Cross
3) Tinamou (partridge-like bird), "coal sack" below Southern Cross
4) Llama, between Southern Cross and epsilon-Scorpio
5) Baby Llama, "below" mother Llama
6) Fox, between tail of Scorpio and Sagittarius

7) a second Tinamou, in Scutum

These dark constellations – in particular, the Llama and the Fox - are certainly the same identified by the Inkas more than six hundreds years ago.[8] In Urton's book however a further possible constellation appears:

"Name: Choque-chinkay , Translation: "golden cat", Provenience: Sonqo, Identification: Tail of Scorpio (or dark spot inside tail?)" (Urton 1982, entry 19, table 7, pag. 99).

The *Choque-chinkay* certainly belongs to the Inka lore of the sky as well, because it appears in the diagram depicted by Pachacuti Yamqui in 1613 (Fig. 5). In this diagram the author represents the objects which, according to him, were venerated by the Inka in the Coricancha. I will not describe it in much details neither go trough the problem of its interpretation, but only notice some of its elements.[9] The space is vertically divided, and it contains stellar objects (for instance, Orion on the upper left, the Southern cross at the centre, seven Pleiades) solar system objects (the Sun, the Moon, Venus as morning and evening star) meteorological objects (the rainbow) natural "earthly" objects (the surface of the earth, the sacred tree).

There is only *one* element which looks symbolic and does not allow an immediate interpretation. It is the stylised figure on the middle right which is called Choque-chinkay and usually described as a "crying" feline. There is no doubt that "crying" figures like this are associated with the rain and, in particular, with the dark clouds promising a storm.[10] However, a dual interpretation as a dark cloud constellation is not excluded by this. It has, in fact, been shown by Urton that the observation of the "darkness of the dark clouds" was used (and is still in use) to make predictions on the quantity of rain and therefore on the outcome of the seeding. But, where in the sky might this constellation be located? It is in fact difficult to accommodate a further dark constellation in the dark region inside the tail of Scorpio, already in part occupied by the Fox, as tentatively suggested in Urton's book.

To try to understand where the Puma could be located, I will resort again to the sacred landscape of Cusco.

There are many well known examples of ancient cultures which constructed links between "heaven-sky" and "human world-earth" by means of astronomically related buildings. Some such examples are controversial and not all scholars accept them, such as, for instance, the theory which interprets the disposition of the three Giza main pyramids as a representation of the three stars of Orion's belt,[11] but other are certain. The connection between earth and sky was frequently obtained using *hyerophanies* , "sacred machines" which were activated by specific celestial events; among them, the famous *Castillo* of Chichen Itza', in the Yucatan, a toltec-maya pyramid which was constructed in such a way that a light and shadow serpent descends its staircase at the equinox.[12] The key example for us here is however that of the megalithic temples of Malta. These huge buildings, constructed between 3500 and 2500 BC, were planned according to a complex cosmographic concept, which included the "shape" of the so called "mother goddess" (a feminine "fat" deity very probably worshipped there) in the internal layout of the temples, the orientation of the main axis to the rising of the Southern Cross-Centaurus asterism[13] and probably also the orientation of the left "altar" of the temple to the winter solstice sunrise.[14] The Malta temple was therefore a terrestrial image of the goddess, criss-crossed by astronomically oriented directions.

If the Inka, as it looks reasonable, identified a Puma dark cloud constellation, perhaps the Cusco layout was similarly meant as a *replica* of the "celestial puma" (according to a old tradition which survives nowadays, the fortress of Paramonga could have been planned in the shape of a Llama, perhaps an image of the celestial one as well). But Cusco lies at the confluence of two rivers in the Vilcanota, and thus one is tempted to suppose that the celestial Puma should share the same property. Actually, all the dark cloud constellations identified by Urton are associated only with that "southern" part of the Milky Way which connects Scutum with Canis Major. This is the part of our galaxy which displays the brightest star luminosity and, as a consequence, a sharp contrast with the dark zones. It forms, at Cusco, a complete arc in the sky around midnight at the autumnal equinox, which is therefore the best period for viewing it. There is, however, a connection, shown by Urton, of the dark clouds with the rainy season (October to April), and the "northern" part of the Milky Way is clearly visible, in the beginning of the rainy season (October-November) divided into two branches, up to our Cygnus constellation where the two branches converge. According to one of Urton's informants:

"The Milky Way, he said, is actually made up of two rivers, not one. The two Mayus originate at a common point in the north, flow in opposite directions from north to south, and collide head on in the in the southern Milky Way…These data indicate that the celestial river has a second center, a "center of origin", in the north."
 (Urton 1982, page 59).

I am, therefore, proposing that the Puma dark cloud constellation might be located in this "center", thus between Cygnus and Vulpecula, exactly as Cusco was located at the "navel" of the world and at the confluence of two rivers.
Perhaps further anthropological research could be of help in clarifying existence and position of the Puma constellation. As an overall observation, one can note that also "standard" (i.e. made out of lines between stars) "northern" constellations such as Ursa Major and Cygnus are absent from the descriptions of Urton's informants (only the bright star Deneb appears as a "marker of the northern quarter"). However, for instance, there is an unidentified constellation (called *Passon Cruz,* entry 29 in Urton's catalogue) which resembles Cygnus.

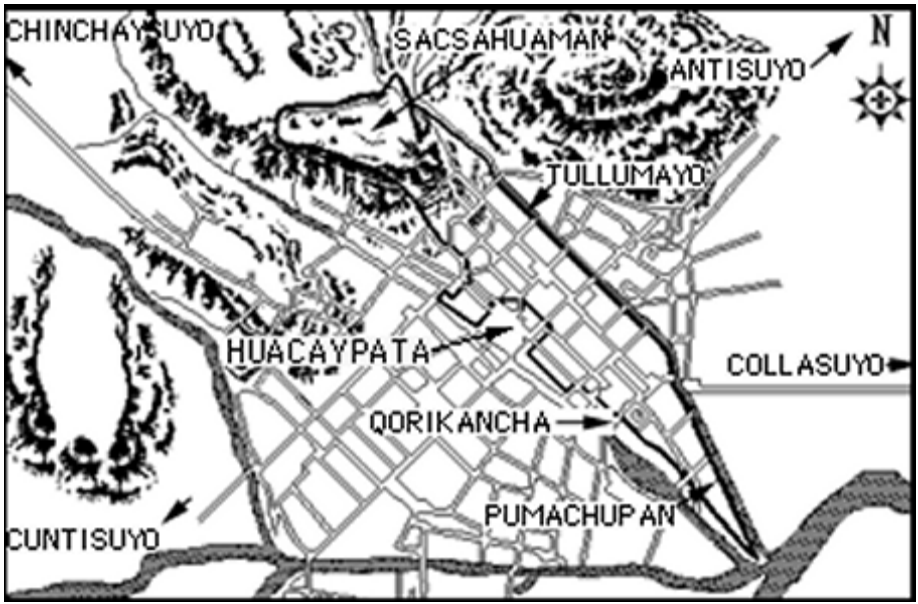

Fig. 1 The ancient part of Cusco in the shape of a puma.

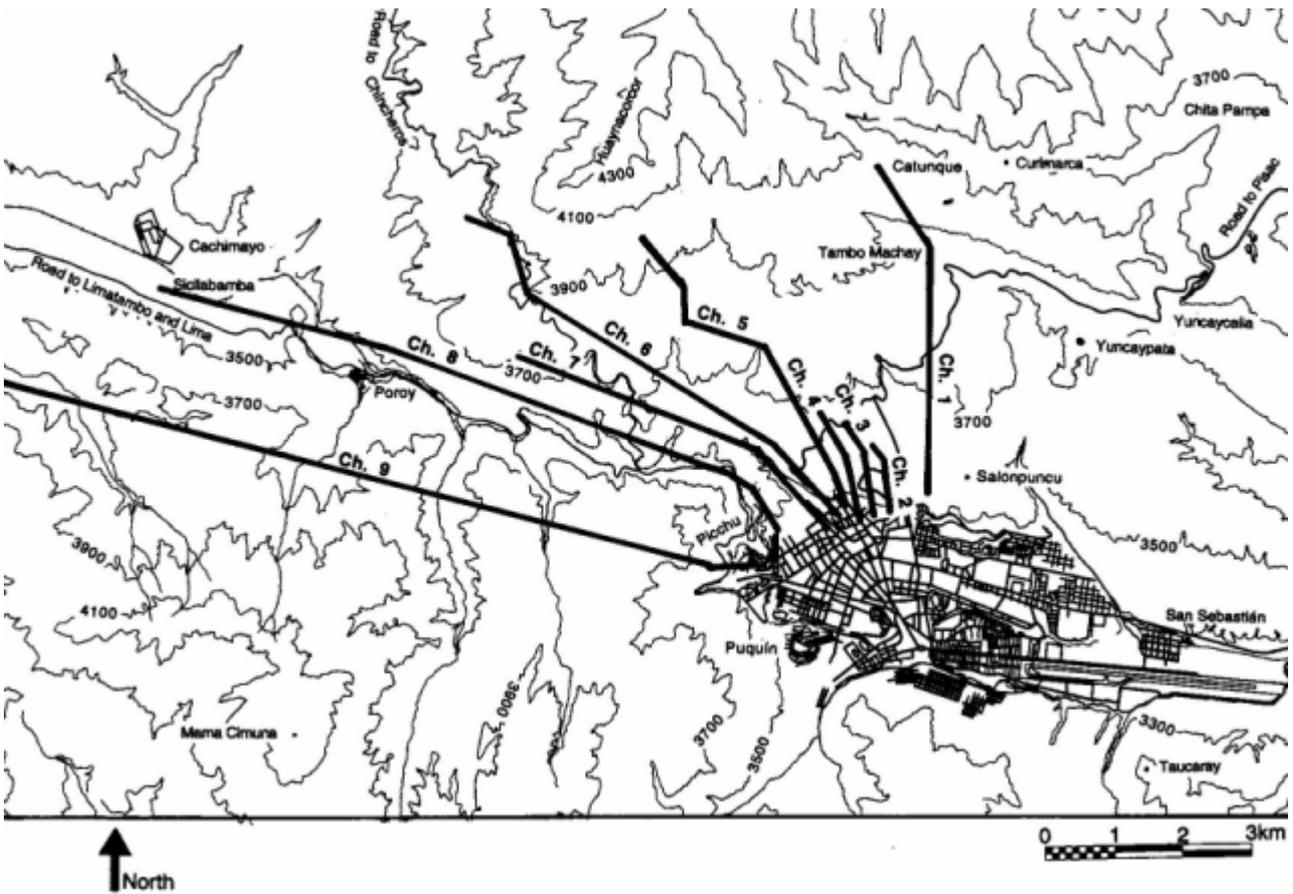

Fig. 2 An example of the structure of the ceque system: the first nine ceques of *Antisuyu* in Bauer's reconstruction (source: Bauer 1998).

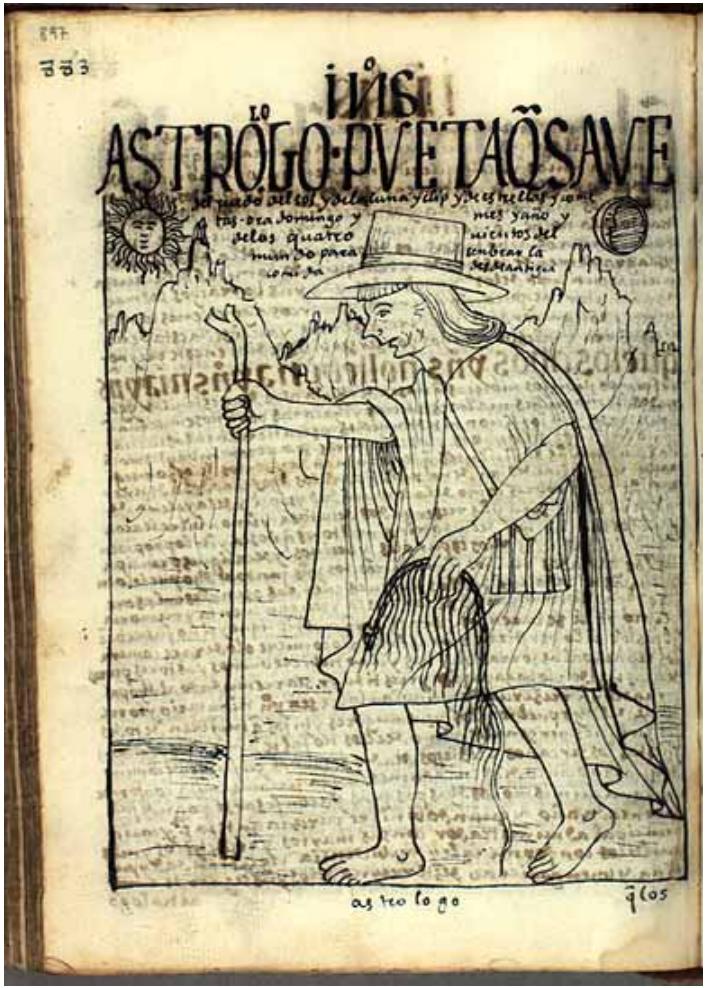

Fig. 3 The Inka astrologer, from *Nuova Cronica y Buen Gobierno* by Guaman Poma de Ayala, carrying a Khipu.

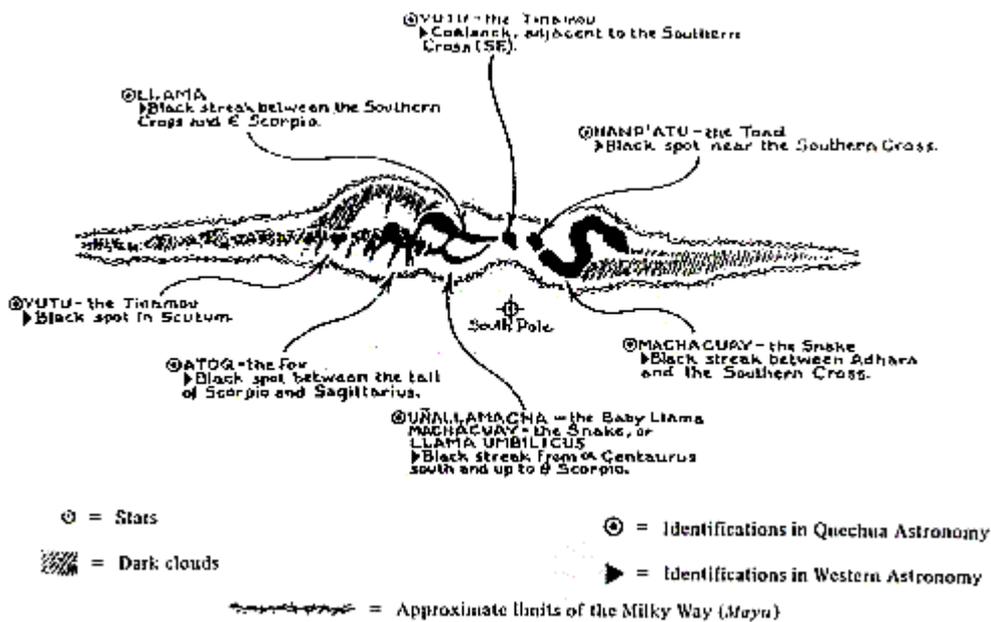

Fig. 4 The dark cloud constellations identified by Urton (Source: Urton 1982).

Fig. 5 Diagram depicted by Pachacuti Yamqui in 1613

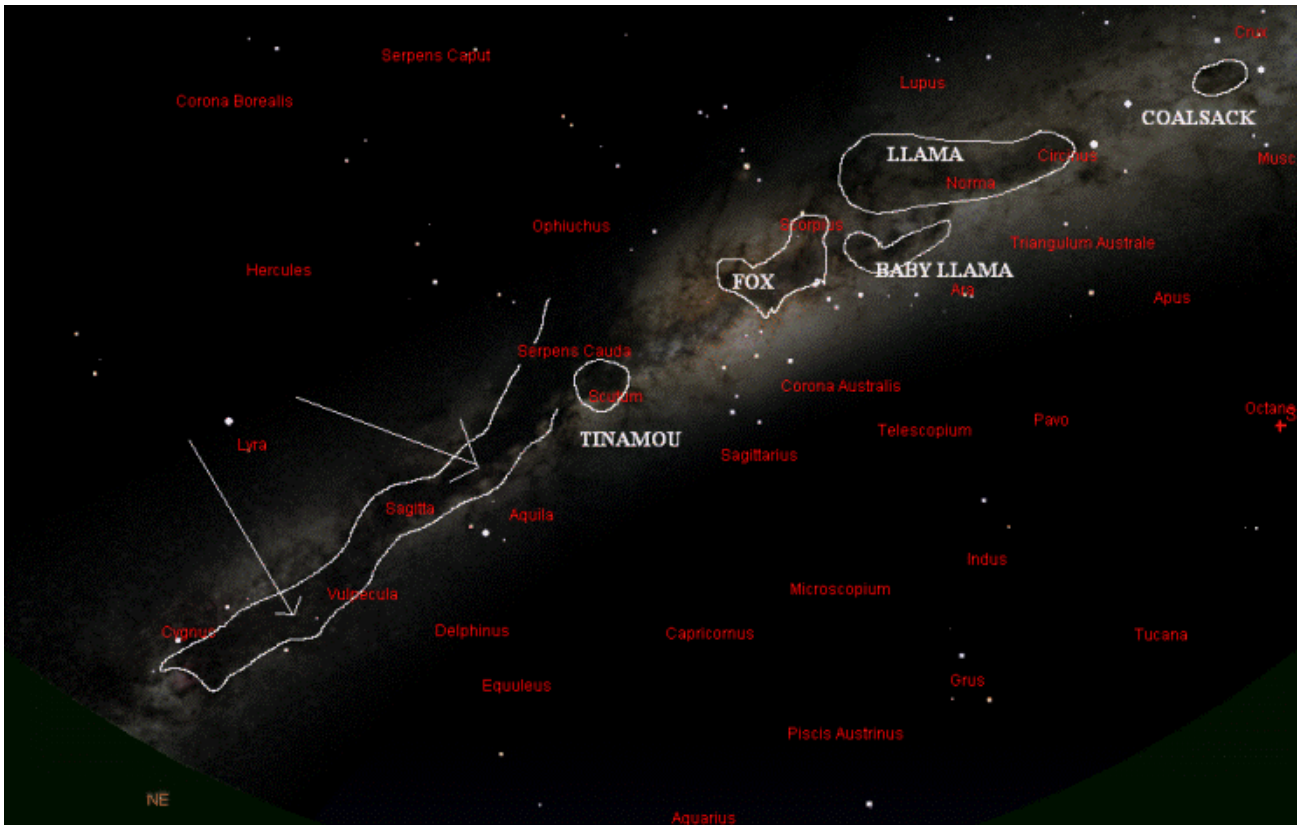

Fig. 6. The Milky Way in the region between Cygnus and Crux, as viewed from Cusco in 1400 BC. The dark cloud constellations identified by Urton in this region have been sketched as an help to the eye. The huge region "between two rivers" which is comprised in the band connecting Scutum and Cygnus is indicated by arrows.